\documentclass[twocolumn,aps,floats,showpacs,superscriptaddress]{revtex4}
\usepackage{amsmath}

\usepackage{epsfig}
\usepackage{graphicx}
\usepackage{dcolumn}
\usepackage{bm}

\def\gapp{\lower.35em\hbox{$\stackrel{\textstyle>}{\sim}$}}
\def\lapp{\lower.35em\hbox{$\stackrel{\textstyle<}{\sim}$}}

\begin{document}
\bibliographystyle{apsrev}
%

\title{
Interactions, disorder and local defects in graphite}
\author{M.A.H. Vozmediano,}
\affiliation{
Departamento de Matem\'aticas, Unidad Asociada
CSIC-UC3M, Universidad
Carlos III de Madrid, E-28911 Legan\'es, Madrid, Spain.}
\author{F. Guinea$^2$, and M. P. L\'opez-Sancho}
\affiliation{
Instituto de Ciencia de Materiales de Madrid,
CSIC, Cantoblanco, E-28049 Madrid, Spain. }
\date{\today}
\begin{abstract}
Recent experiments report the existence of ferromagnetic
and superconducting fluctuations in graphite at unexpectedly
high temperatures.  The interplay of disorder and interactions
in a 2D graphene layer is shown to give rise to a rich phase
diagram where strong coupling phases can become stable.
Local defects can explain the ferromagnetic signals.
\end{abstract}
%
\pacs{75.10.Jm, 75.10.Lp, 75.30.Ds, 71.20.Tx, 73.50.Gr}
%
%
\maketitle
{\it Introduction.}
A number of recent experiments suggest that pure graphite
behaves as a highly correlated electron system\cite{esquireview}.
In particular it shows a metal-insulator transition in magnetic fields and
insulating behavior in the direction perpendicular to the planes in different
samples\cite{Ketal00,Eetal02,KEK02,Ketal02,MHM02,Cetal02,Ketal03,Ketal03b}.
The interest in this material is focussed nowadays in the observation
of ferromagnetic behavior\cite{Hetal03},
enhanced by proton bombardment\cite{proton} what opens a new way to the
creation of organic magnets.

In refs.\cite{Gon93,Gon94}
a simple microscopic model was proposed as a new framework to
study the  physics of 2D graphene sheets and its topological
variant fullerenes and carbon nanotubes. The main
assumption of the model is to neglect the
coupling between layers and consider graphite as
a pure two-dimensional system. This assumption is supported
by experiments where an anisotropy of up to three orders
of magnitude is measured in magnetotransport\cite{Ketal02}.
The model predicts
non-Fermi liquid  behavior for the graphene system
and can account for the linear behavior with energy of the
quasiparticle scattering rate\cite{GGV96} observed in photoemission
experiments\cite{lifexp}.

In this work we review the main features of the model with and
without disorder and
propose a new mechanism to explain the ferromagnetic
fluctuations observed in the experiments.

\begin{figure}
\centerline{\epsfig{file=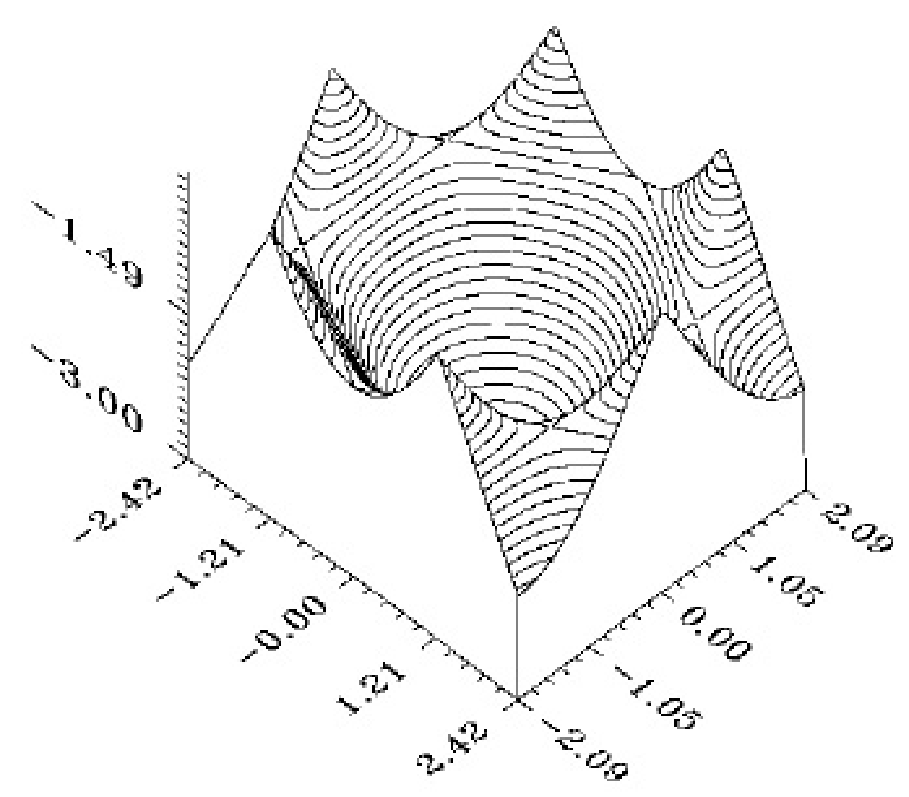,width=3in}}
\caption{Lower branch of the electronic dispersion
relation. The cusps appear at the six corners of the
first Brillouin zone.}
\label{fig1}
\end{figure}

{\it The model. RG results.} The conduction band of graphite is well
described by tight binding models which include only
the $\pi$ orbitals which are perpendicular to the graphite
planes at each C atom\cite{SW58}.
The two dimensional hexagonal lattice of a graphene plane
has two atoms per unit cell. A tight binding calculation
with only nearest neighbors hopping gives rise to the
dispersion relation
\begin{align}
E({\bf k})=\pm t\sqrt{1+4\cos^2\frac{\sqrt{3}}{2}k_x+
4\cos\frac{\sqrt{3}}{2}k_x\cos\frac{3}{2}k_y}
\label{disprel}
\end{align}
whose  lower branch is shown in Fig.\ref{fig1}.
This dispersion relation gives rise at half filling
to a Fermi surface consisting of six isolated points
two of which are inequivalent.
A low-energy effective Hamiltonian can be defined by
expanding the dispersion relation about any of the Fermi points.
The resulting Hamiltonian has the form of
a massless two dimensional Dirac Hamiltonian.
The Fermi velocity, $v_F$, can be expressed in terms of the
matrix elements between nearest neighbor $\pi$ orbitals, $t$,
as $v_F = (3 t a)/2 $, where $a$ is the C-C distance.

The electronic states within each graphene
plane are described by two two-component spinors associated to the two
inequivalent Fermi points in the Brillouin Zone.
The Hamiltonian of the free system is:
\begin{align}
H_0&=iv_F\int d^2x \overline\Psi(\vec x)\vec\gamma
\cdot\vec\nabla\Psi(\vec x)
\label{hamil}
\end{align}
where the two-dimensional $\gamma$ matrices are built
as appropriate combinations of Pauli matrices\cite{Gon93}.
The Hamiltonian (\ref{hamil}) gives an effective description of graphite
in an energy range bound by a lower cutoff $\sim .27 eV$
dictated by the interlayer coupling,
and a higher cutoff, where the bands can no longer be
approximated by a linear dispersion relation $\sim 3-4 eV$.

The Hamiltonian (\ref{hamil}) is the perfect model
for Renormalization Group (RG) calculations.
It is scale invariant and does not have the complications
of an extended Fermi surface. The model is similar to
the $D=1$ electron system in that it has Fermi points and linear
dispersion. Nevertheless
naive dimensional analysis shows that
four and more Fermi interactions are irrelevant in this two-dimensional
case. The only interaction that may survive at low energies
is the long (infinite) range Coulomb interaction,
unscreened because of the vanishing density of states at
the Fermi point. Following the quantum field theory nature of
the model, we trade the classical Coulomb interaction
\begin{align}
H_{ee}&=\frac{v_F}{4\pi}\int d^2x d^2x'\overline\Psi(\vec x)
\gamma_0\Psi(\vec x)\frac{g}{|\vec x-\vec x'|}
\overline\Psi(\vec x')\gamma_0\Psi(\vec x')
\end{align}
where $g=e^2/4\pi v_F$ is the dimensionless coupling constant,
by a local gauge interaction through a minimal coupling.
\begin{equation}
L_{int}=g\int d^2x dt j^\mu (x,t)A_\mu (x,t)\;,
\end{equation}
where the electron current is defined as
$$j^\mu =(\overline\Psi\gamma^0\Psi,v_F \overline\Psi\gamma^i\Psi)\;.$$
This interaction is marginal in the RG sense,
all the rest are irrelevant. The RG analysis of the model
gives the following results \cite{Gon93,Gon94}:

1. From the computation of the electron self-energy at the one
loop level we get a non trivial renormalization
of the Fermi velocity that {\bf grows in the infrared}. This
result implies a breakdown of the relation between the energy
and momentum scaling, a signature of a quantum critical point.

2. The electron-photon vertex and the photon propagator
are not renormalized at the one loop level. This means
that the electric charge is not renormalized, a
result that could be predicted by gauge invariance,
and it also implies that the effective coupling constant
$g=e^2/4\pi v_F$ {\bf decreases at low energies} defining
an infrared free fixed point of the RG. It is interesting
to note that the Lorentz invariance of the model that
was explicitly broken by the Fermi velocity is
recovered at the fixed point since the
velocity  of light, c, fixes
a limit to the growing of the Fermi velocity.

2. From the electron self-energy at two loops order we get
a non trivial wave function renormalization meaning that
the infrared stable fixed point corresponds to a free fixed
point different from the Fermi liquid. This result has been
shown to persist
in the  non-perturbative regime\cite{Gon99}. This is a non-trivial
result that has physical implications. In particular
it implies that the inverse quasiparticle lifetime
increases linearly with energy\cite{GGV96}, a result that has been
observed experimentally in \cite{lifexp} in the energy range
of validity of the model.

In conclusion, we have shown that
without disorder, edges, or other
perturbations, the graphene system at low energies
has gapless excitations differing from the Fermi
liquid quasiparticles but does not support
magnetic or superconducting instabilities.

The strong coupling regime of the graphene
system has been analyzed in \cite{Khv01,Khv01b}.
There it is argued that a dynamical breakdown of the chiral
symmetry (degeneracy between the two Fermi points) will
occur at strong coupling and a gap will open in the spectrum
forming a kind of charge density wave. Graphite can
then be seen as an excitonic insulator that can
become ferromagnetic upon doping. Being
non-perturbative, these phenomena are likely to
be washed out by any amount of disorder at intermediate
energies.

{\it Inclusion of disorder.}

The previous description analyzes
the small momentum scattering
due to the long range Coulomb interaction, as it is the only one which leads to
logarithmically divergent perturbative corrections. Some electronic
instabilities, like ferromagnetism or
anisotropic superconductivity, require
the existence of short range interactions with significant
strength. The irrelevant character
of short range interactions can be changed by the presence
of disorder that enhances the density of states at
the Fermi level.

Disorder can be included in the renormalization
group scheme by the introduction of random gauge
fields. This is a standard procedure in the study of
the states described by the two dimensional Dirac
equation associated to random lattices
or to integer quantum Hall transitions\cite{CMW96,Cetal97,HD02}.
There it is seen that, usually, the density of
states at low energies is increased. To demonstrate
how these gauge fields can arise in the graphene system,
we will describe in detail a special type of disorder
that we call topological disorder.

\begin{figure}
\centerline{\epsfig{file=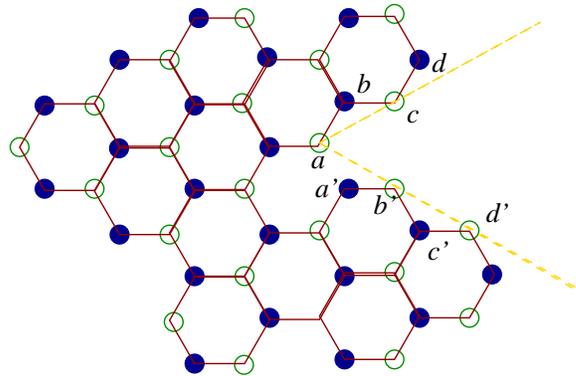,width=3in}}
\caption{Formation of a pentagonal ring in the honeycomb
lattice. Points $a, b, c, d ...$ have to be identified
with points $a', b', c', d' ...$. The defect can be seen
as a disclination, defined by the straight dashed lines.}
\label{fig5}
\end{figure}

The formation of pentagons and heptagons in the lattice, without affecting
the threefold coordination of the carbon atoms, lead to the warping of
the graphene sheets, and are responsible for the formation of curved
fullerenes, like C$_{60}$. They can be viewed as disclinations
in the lattice, and, when circling one such defect, the
two sublattices in the honeycomb structure are exchanged
(see Fig.[\ref{fig5}]).
The two fermion flavors associated to the two
Fermi points are also exchanged
when moving around such a defect. The
scheme to incorporate this change in a continuum
description was discussed in\cite{Gon99}.
The process can be described by means
of a non Abelian gauge field, which rotates the spinors in
flavor space. The vector potential is that of a vortex at the
position of the defect, and the flux is $\pm \pi / 2$.

Dislocations can be analyzed in terms of bound disclinations,
that is, a pentagon and an heptagon located at short distances,
which define the Burgers vector of the dislocation. Thus,
the effect of a dislocation on the electronic levels of
a graphene sheet is analogous to that of the vector potential
arising from a vortex-antivortex pair. We can extend this description\cite{G98},
and assume that a lattice distortion which rotates the lattice axis
can be parametrized by the angle of rotation, $\theta (
{\bf \vec{r}} )$,  of the local axes
with respect to a fixed reference frame. Then, this distortion
induces a gauge field such that:
\begin{equation}
 {\bf \vec{A}} ( {\bf \vec{r}} )  = 3 \nabla \theta ( {\bf \vec{r}} )
\left( \begin{array}{cc} 0 &-i \\ i &0 \end{array} \right)
\label{gauge}
\end{equation}
Thus, a random distribution of topological defects can be described
by a (non abelian) random gauge field.

Other types of disorder can similarly be associated to random
gauge fields. The complete Hamiltonian of the system can be written as
\begin{align}
H=H_{ee}+H_{disorder}
\end{align}
where
\begin{align}
H_{ee}&=\frac{v_F}{4\pi}\int d^2x d^2x'
\overline\Psi(\vec x)\gamma_0\Psi(\vec x)\frac{g}{|\vec x-\vec x'|}
\overline\Psi(\vec x')\gamma_0\Psi(\vec x')
\end{align}
\begin{align}
\label{Hdisorder}
H_{disorder}=\frac{v_\Gamma}{4}\int d^2x
\overline\Psi(\vec x)\Gamma\Psi(\vec x)A(\vec x)
\end{align}
$v_\Gamma$ characterizes the strength and the
$4\times4$ matrix $\Gamma$ the type of the vertex.
In general, $A ( \vec x )$ is a quenched, Gaussian variable with the
dimensionless variance $\Delta$, i.e.,
\begin{align}
\label{Gauss}
\langle A(\vec x)\rangle=0\quad,\quad\langle
A(\vec x)A(\vec x')\rangle=\Delta\delta^2(\vec x-\vec x')\quad.
\end{align}
In ref. \cite{SGV03} a complete RG study of the disordered
system was analyzed by adding gauge couplings associated to all possible
gamma matrices.
i) For a random chemical potential, the $4\times4$
matrix $\Gamma$ is given by $\Gamma=\gamma_0$.
The long range components of this
 type of disorder do not induce transitions between the
two inequivalent Fermi points.
This type of disorder yields an unstable fixed line.
ii) A random gauge potential involves the $4\times4$ matrices
$\Gamma=i\gamma_1$ and $\Gamma=i\gamma_2$.
This type of disorder  gives rise to a stable fixed line which
is linear in the $(g,\Delta)$-plane.
iii) (a) A fluctuating mass term is described by
$\Gamma={\bf{1}}_{4\times4}$.
(b) Topological disorder is given by $\Gamma=i\gamma_5$ with
$\gamma_5={\bf{1}}_{2\times2}\otimes\sigma_2$.
c) To complete the discussion, we also mention
$\Gamma=i\tilde\gamma_5$
where $\tilde\gamma_5={\bf{1}}_{2\times2}\otimes\sigma_1$.
This vertex type
can be related to an imaginary mass that couples the
two inequivalent Fermi points.
All these types of disorder will yield a stable
fixed line which is cubic in the $(g,\Delta)$-plane.

The phase diagram obtained in  \cite{SGV03} is reproduced in Fig. 3.
\begin{figure}

       \epsfig{file=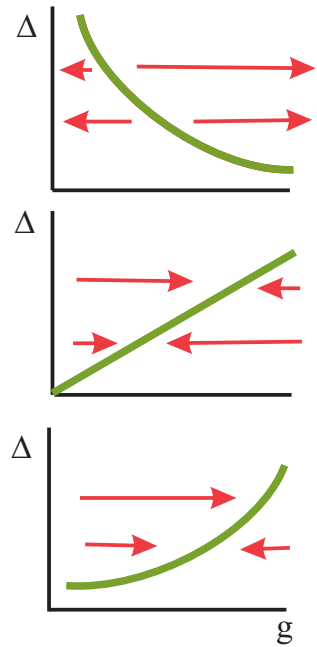,height=8.5cm}
    \caption{One-loop phase diagram for two-dimensional massless Dirac spinors
including long-ranged electron-electron interaction $g$ and
disorder $\Delta$. Top: Random chemical potential
($\Gamma=\gamma_0$). Center: Random gauge potential
($\Gamma=i\gamma_1,i\gamma_2$). Bottom: Random mass term
($\Gamma={\bf{1}}_{4\times4}$), topological disorder
($\Gamma=i\gamma_5$), and $\Gamma=i\tilde\gamma_5$.}
    \label{Phasediagram}
\end{figure}

 i) For a random chemical potential $(\Gamma=\gamma_0)$,
$v_\Gamma=v_1$ remains constant under
renormalization group transformation. There is  an unstable fixed
line at $v_F^*=v_1^2\Delta/(2e^2)$. In the $(g,\Delta)$-plane,
the strong-coupling and the weak-coupling phases are separated
by a hyperbola, with the critical electron interaction
$g^*=e^2/v_F^*=2e^4/(v_1^2\Delta)$.
ii) A random gauge potential involves the vertices $\Gamma=i\gamma_1,i\gamma_2$.
The vertex strength renormalizes as $v_\Gamma=v_F$.  There is thus an
attractive Luttinger-like fixed point for each disorder correlation strength
$\Delta$ given by $v_F^*=2e^2/\Delta$ or $g^*=\Delta/2$.
iii) For a random mass term $\Gamma={\bf{1}}_{4\times4}$, topological disorder
$\Gamma=i\gamma_5$, and $\Gamma=i\tilde\gamma_5$, we have $v_\Gamma=v_F^2/v_3$.
There is thus again an attractive Luttinger-like fixed point for each
disorder correlation strength $\Delta$ given by $v_F^*=\root 3\of{2v_3^2e^2/\Delta}$
or $g^*=\root3\of{\Delta e^4/(2v_3^2)}$.

The most interesting phase is the one induced by
a random gauge potential,  a random mass term or the
topological disorder. All of them
drive the system towards a new stable,
Luttinger-like fixed point.
This phase is characterized by a vanishing quasiparticle residue,
leading to anomalous one particle properties. The Luttinger liquid
features associated to this fixed line are notoriously difficult
to observe, although they can be probed in tunneling experiments,
or by measuring the peak width in ARPES. They will also influence
the interlayer transport properties\cite{VLG02,VLG03}. Small
perturbations by other types of disorder, like a random local
potential induce a flow along this fixed line, as in the absence
of interactions\cite{LFSG94}. The strong coupling fixed point describes, most
likely, a disordered insulating system.

{\it Localized states.}
In addition to the extended disorder discussed previously, a
graphene plane can show states localized at
interfaces\cite{WS00,W01}, which, in the absence of other types of
disorder, lie at the Fermi energy. Changes in the local
coordination can also lead to localized states\cite{OS91}.

The tight binding model defined by the $\pi$ orbitals at the
lattice sites can have edge states when the sites at the edge
belong all to the same sublattice(zig-zag edges)\cite{WS00,W01}.
These states lie at zero energy, which, for neutral graphene planes, correspond
to the Fermi energy.

\begin{figure}[h]
  \begin{center}
    \epsfig{file=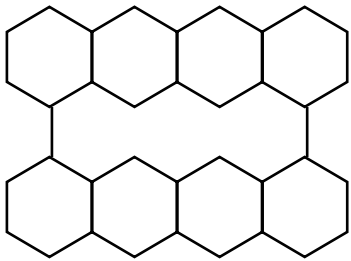,height=6cm}
    \caption{Example of a crack in a graphene plane.
The atoms at the upper edge and those at the lower edge belong to
different sublattices. }
    \label{graphite_crack}
\end{center}
\end{figure}

In a strongly disordered sample, large defects made up of many vacancies
can exist. These defects give rise to localized states, when the termination
at the edges is locally similar to the surfaces discussed above\cite{GLSV04}. Note that,
if the bonds at the edges are saturated by bonding to other elements, like
hydrogen, the states at these sites are removed from the Fermi
energy, but a similar boundary problem arises for the remaining
$\pi$ orbitals. A particular simple example is given by the crack
shown in Fig.[\ref{graphite_crack}].

These states are half filled in a neutral graphene plane.
In the absence of electron interactions, this leads
to a large degeneracy in the ground state. A finite local repulsion
will tend to induce a ferromagnetic alignment of the electrons
occupying these states, as in similar cases with degenerate
bands\cite{Vetal99}. Hence, we can assume that the presence of
these states leads to magnetic moments localized near the
defects.

We now have to analyze the influence of
 these magnetic moments in conduction band described in the previous sections.
The hopping between the states involved in the formation of these moments
and the delocalized states in the conduction band vanishes by definition,
if the localized states lie at zero energy. Hence, a Kondo like coupling
mediated by the hopping will not be induced. The localized and
conduction band states, on the other hand, are defined on the same lattice sites.
The existence of a finite local repulsion, $U$, will lead to an effective
ferromagnetic coupling. The local coupling, at site $i$,
between the localized states and the
conduction band is proportional to $U \sum_j \rho_{i,j}$, where
$\rho_{i,j}$ is charge of state $j$ at site $i$.
In order to get an order of magnitude estimate of the effect
of these states, we will assume that the number of states
induced near a vacancy is similar to the number of atoms
at its edge, ${\cal N}$, and that these states are sufficiently
localized around the vacancy. Hence, each vacancy nucleates
a moment of order ${\cal N}$. The effective coupling between
a vacancy and the conduction electrons is proportional to
$U {\cal N}$, and it is distributed over an area $\sim {\cal N}^2$.
The conduction electrons will mediate an RKKY interaction between
the localized moments:
\begin{equation}
J_{RKKY} ( {\bf \vec{r}} ) \sim
U^2  \int
d^2 {\bf k} e^{i {\bf \vec{k}} {\bf \vec{r}}
} \chi ( {\bf \vec{k}} ) \sim U^2 \frac{a^4}{v_F | {\bf \vec{r}} |^3}
\label{RKKY}
\end{equation}
 Where the static
susceptibility is
 $\chi ( {\bf \vec{k}} ) \propto | {\bf \vec{k}} |$\cite{GGV96},
and $a$ is the lattice constant.
It is interesting to note that, due to
the absence of a finite Fermi surface, the RKKY interaction
in eq.(\ref{RKKY}) does not have oscillations. Hence, there are no
competing ferro- and antiferromagnetic couplings, and the
magnetic moments will tend to be ferromagnetically aligned,
leading to an effective magnetic field, $H_{ext} ( {\bf \vec{r}} )$,
with non zero
average, acting on the conducting electrons.

From power counting, this coupling is relevant in the
Renormalization Group sense. Thus, in the presence of
extended vacancies, the RG flow discussed in the previous
sections has to be arrested at scales comparable
to $\langle H_{ext} ( {\bf \vec{r}} ) \rangle
\sim  U {\cal N}  \rho_{vac}$, where $\rho_{vac}$ is the concentration
of the large vacancies which may give rise to localized
states. At lower energies, or temperatures, the  graphene
planes with extended vacancies will behave as a ferromagnet.

{\it Conclusions.}
In this work we present a microscopic model for
studying the low energy
properties of a single graphene
layer  as a model relevant for some graphite samples showing
two-dimensional anomalous behavior. In particular we tried
to envisage a model able to explain the ferromagnetism
observed recently in a variety of graphitic materials.

The model is based on the particular dispersion relation of the
2D honeycomb lattice that, at half filling, has Fermi points instead
of Fermi lines. The linearization of the dispersion about a
Fermi point gives rise to a model similar to the one-dimensional
electron system with zero density of states at the Fermi level.
Unlike the 1D case, the real two-dimensional nature of the
present model makes the four Fermi interactions irrelevant
in the renormalization group sense
while the long range Coulomb interaction is unscreened and plays
an important role. It renormalizes the Fermi velocity that
grows at low energies while the effective charge is
not renormalized, a consequence of the gauge invariance.
The effective coupling constant $e^2/v_F$ goes to zero
 driving the system to a non-trivial infrared free
fixed point.
As a consequence of the singular Coulomb interaction,
the electron acquires anomalous
dimension and the quasiparticle scattering rate grows
linearly with frequency at intermediate
frequencies, as observed in experiments. The model does not
support magnetic or any other short range interactions
at this level.

The presence of disorder changes the previous situation in various
respects. We have considered two types of disorder, non-local
disorder as the one produced by
impurities or lattice distortions, modelled by the coupling
of the electrons to random gauge fields as
in the random lattice models, and local large defects
as the ones produced
in the experiments by proton bombardment.
Extended disorder gives rise to a rich phase diagram
with strong coupled phases whose physical properties are
still to be analyzed. Local defects  give
rise to the appearance of local moments whose interaction can
induce ferromagnetism in large portion of the sample.

We thank P. Esquinazi for sharing his experiments
with us and for many illuminating discussions.
Funding from MCyT (Spain) through grant MAT2002-0495-C02-01 is
acknowledged.
\bibliography{sns04}

\newcommand{\npb}{Nucl. Phys.}\newcommand{\adv}{Adv.
  Phys.}\newcommand{\epl}{Europhys. Lett.}
\begin{thebibliography}{10}
\expandafter\ifx\csname bibnamefont\endcsname\relax
  \def\bibnamefont#1{#1}\fi
\expandafter\ifx\csname bibfnamefont\endcsname\relax
  \def\bibfnamefont#1{#1}\fi
\expandafter\ifx\csname url\endcsname\relax
  \def\url#1{\texttt{#1}}\fi
\expandafter\ifx\csname urlprefix\endcsname\relax\def\urlprefix{URL }\fi
\providecommand{\bibinfo}[2]{#2}
\providecommand{\eprint}[2][]{\url{#2}}

\bibitem{esquireview}
\bibinfo{author}{\bibfnamefont{P.}~\bibnamefont{Esquinazi}} \bibnamefont{and}
  \bibinfo{author}{\bibnamefont{et~al}}, \bibinfo{journal}{Advances in Solid
  State Physics} \textbf{\bibinfo{volume}{43}}, \bibinfo{pages}{207}
  (\bibinfo{year}{2003}).

\bibitem{Ketal00}
\bibinfo{author}{\bibfnamefont{Y.}~\bibnamefont{Kopelevich}},
  \bibinfo{author}{\bibfnamefont{P.}~\bibnamefont{Esquinazi}},
  \bibinfo{author}{\bibfnamefont{J.~H.~S.} \bibnamefont{Torres}},
  \bibnamefont{and}
  \bibinfo{author}{\bibfnamefont{S.}~\bibnamefont{Moehlecke}},
  \bibinfo{journal}{J. Low Temp. Phys.} \textbf{\bibinfo{volume}{119}},
  \bibinfo{pages}{691} (\bibinfo{year}{2000}).

\bibitem{Eetal02}
\bibinfo{author}{\bibfnamefont{P.}~\bibnamefont{Esquinazi}},
  \bibinfo{author}{\bibfnamefont{A.}~\bibnamefont{Setzer}},
  \bibinfo{author}{\bibfnamefont{R.}~\bibnamefont{H\"ohne}},
  \bibinfo{author}{\bibfnamefont{C.}~\bibnamefont{Semmelhack}},
  \bibinfo{author}{\bibfnamefont{Y.}~\bibnamefont{Kopelevich}},
  \bibinfo{author}{\bibfnamefont{D.}~\bibnamefont{Spemann}},
  \bibinfo{author}{\bibfnamefont{T.}~\bibnamefont{Butz}},
  \bibinfo{author}{\bibfnamefont{B.}~\bibnamefont{Kohlstrunk}},
  \bibnamefont{and} \bibinfo{author}{\bibfnamefont{M.}~\bibnamefont{L\"osche}},
  \bibinfo{journal}{\prb} \textbf{\bibinfo{volume}{66}},
  \bibinfo{pages}{024429} (\bibinfo{year}{2002}).

\bibitem{KEK02}
\bibinfo{author}{\bibfnamefont{H.}~\bibnamefont{Kempa}},
  \bibinfo{author}{\bibfnamefont{P.}~\bibnamefont{Esquinazi}},
  \bibnamefont{and}
  \bibinfo{author}{\bibfnamefont{Y.}~\bibnamefont{Kopelevich}},
  \bibinfo{journal}{\prb} \textbf{\bibinfo{volume}{65}},
  \bibinfo{pages}{241101} (\bibinfo{year}{2002}).

\bibitem{Ketal02}
\bibinfo{author}{\bibfnamefont{Y.}~\bibnamefont{Kopelevich}},
  \bibinfo{author}{\bibfnamefont{P.}~\bibnamefont{Esquinazi}},
  \bibinfo{author}{\bibfnamefont{J.~H.~S.} \bibnamefont{Torres}},
  \bibinfo{author}{\bibfnamefont{R.~R.} \bibnamefont{da~Silva}},
  \bibnamefont{and} \bibinfo{author}{\bibfnamefont{H.}~\bibnamefont{Kempa}}, in
  \emph{\bibinfo{booktitle}{Studies of High Temperature Superconductors}},
  edited by \bibinfo{editor}{\bibfnamefont{A.}~\bibnamefont{Narlikar}}
  (\bibinfo{publisher}{Nova Science Pub., Inc}, \bibinfo{year}{2003}),
  vol.~\bibinfo{volume}{45}, p.~\bibinfo{pages}{59}.

\bibitem{MHM02}
\bibinfo{author}{\bibfnamefont{S.}~\bibnamefont{Moehlecke}},
  \bibinfo{author}{\bibfnamefont{P.-C.} \bibnamefont{Ho}}, \bibnamefont{and}
  \bibinfo{author}{\bibfnamefont{M.~B.} \bibnamefont{Maple}},
  \bibinfo{journal}{Phil. Mag. Lett} \textbf{\bibinfo{volume}{82}},
  \bibinfo{pages}{1335} (\bibinfo{year}{2002}).

\bibitem{Cetal02}
\bibinfo{author}{\bibfnamefont{J.~M.~D.} \bibnamefont{Coey}},
  \bibinfo{author}{\bibfnamefont{M.}~\bibnamefont{Venkatesan}},
  \bibinfo{author}{\bibfnamefont{C.~B.} \bibnamefont{Fitzgerald}},
  \bibinfo{author}{\bibfnamefont{A.~P.} \bibnamefont{Douvalis}},
  \bibnamefont{and} \bibinfo{author}{\bibfnamefont{I.~S.}
  \bibnamefont{Sanders}}, \bibinfo{journal}{Nature}
  \textbf{\bibinfo{volume}{420}}, \bibinfo{pages}{156} (\bibinfo{year}{2002}).

\bibitem{Ketal03}
\bibinfo{author}{\bibfnamefont{Y.}~\bibnamefont{Kopelevich}},
  \bibinfo{author}{\bibfnamefont{J.~H.~S.} \bibnamefont{Torres}},
  \bibinfo{author}{\bibfnamefont{R.~R.} \bibnamefont{da~Silva}},
  \bibinfo{author}{\bibfnamefont{F.}~\bibnamefont{Mrowka}},
  \bibinfo{author}{\bibfnamefont{H.}~\bibnamefont{Kempa}}, \bibnamefont{and}
  \bibinfo{author}{\bibfnamefont{P.}~\bibnamefont{Esquinazi}},
  \bibinfo{journal}{\prl} \textbf{\bibinfo{volume}{90}},
  \bibinfo{pages}{156402} (\bibinfo{year}{2003}).

\bibitem{Ketal03b}
\bibinfo{author}{\bibfnamefont{H.}~\bibnamefont{Kempa}},
  \bibinfo{author}{\bibfnamefont{H.~C.} \bibnamefont{Semmelhack}},
  \bibinfo{author}{\bibfnamefont{P.}~\bibnamefont{Esquinazi}},
  \bibnamefont{and}
  \bibinfo{author}{\bibfnamefont{Y.}~\bibnamefont{Kopelevich}},
  \bibinfo{journal}{Solid State Commun.} \textbf{\bibinfo{volume}{125}},
  \bibinfo{pages}{1} (\bibinfo{year}{2003}).

\bibitem{Hetal03}
\bibinfo{author}{\bibfnamefont{K.}~\bibnamefont{h.~Han}},
  \bibinfo{author}{\bibfnamefont{D.}~\bibnamefont{Spemann}},
  \bibinfo{author}{\bibfnamefont{P.}~\bibnamefont{Esquinazi}},
  \bibinfo{author}{\bibfnamefont{R.}~\bibnamefont{H\"ohne}},
  \bibinfo{author}{\bibfnamefont{R.}~\bibnamefont{Riede}}, \bibnamefont{and}
  \bibinfo{author}{\bibfnamefont{T.}~\bibnamefont{Butz}},
  \bibinfo{journal}{Adv. Mat.} \textbf{\bibinfo{volume}{15}},
  \bibinfo{pages}{1719} (\bibinfo{year}{2003}).

\bibitem{proton}
\bibinfo{author}{\bibfnamefont{P.}~\bibnamefont{Esquinazi}},
  \bibinfo{author}{\bibfnamefont{D.}~\bibnamefont{Spemann}},
  \bibinfo{author}{\bibfnamefont{R.}~\bibnamefont{H\"ohne}},
  \bibinfo{author}{\bibfnamefont{A.}~\bibnamefont{Setzer}},
  \bibinfo{author}{\bibfnamefont{K.-H.} \bibnamefont{Han}}, \bibnamefont{and}
  \bibinfo{author}{\bibfnamefont{T.}~\bibnamefont{Butz}},
  \bibinfo{journal}{Phys. Rev. Lett.} \textbf{\bibinfo{volume}{91}},
  \bibinfo{pages}{227201} (\bibinfo{year}{2003}).

\bibitem{Gon93}
\bibinfo{author}{\bibfnamefont{J.}~\bibnamefont{Gonz\'alez}},
  \bibinfo{author}{\bibfnamefont{F.}~\bibnamefont{Guinea}}, \bibnamefont{and}
  \bibinfo{author}{\bibfnamefont{M.~A.~H.} \bibnamefont{Vozmediano}},
  \bibinfo{journal}{Nucl. Phys. B} \textbf{\bibinfo{volume}{406 [FS]}},
  \bibinfo{pages}{771} (\bibinfo{year}{1993}).

\bibitem{Gon94}
\bibinfo{author}{\bibfnamefont{J.}~\bibnamefont{Gonz\'alez}},
  \bibinfo{author}{\bibfnamefont{F.}~\bibnamefont{Guinea}}, \bibnamefont{and}
  \bibinfo{author}{\bibfnamefont{M.~A.~H.} \bibnamefont{Vozmediano}},
  \bibinfo{journal}{Nucl. Phys. B} \textbf{\bibinfo{volume}{424 [FS]}},
  \bibinfo{pages}{595} (\bibinfo{year}{1994}).

\bibitem{GGV96}
\bibinfo{author}{\bibfnamefont{J.}~\bibnamefont{Gonz\'alez}},
  \bibinfo{author}{\bibfnamefont{F.}~\bibnamefont{Guinea}}, \bibnamefont{and}
  \bibinfo{author}{\bibfnamefont{M.~A.~H.} \bibnamefont{Vozmediano}},
  \bibinfo{journal}{\prl} \textbf{\bibinfo{volume}{77}}, \bibinfo{pages}{3589}
  (\bibinfo{year}{1996}).

\bibitem{lifexp}
\bibinfo{author}{\bibfnamefont{S.}~\bibnamefont{Yu}},
  \bibinfo{author}{\bibfnamefont{J.}~\bibnamefont{Cao}},
  \bibinfo{author}{\bibfnamefont{C.~C.} \bibnamefont{Miller}},
  \bibinfo{author}{\bibfnamefont{D.~A.} \bibnamefont{Mantell}},
  \bibinfo{author}{\bibfnamefont{R.~J.~D.} \bibnamefont{Miller}},
  \bibnamefont{and} \bibinfo{author}{\bibfnamefont{Y.}~\bibnamefont{Gao}},
  \bibinfo{journal}{\prl} \textbf{\bibinfo{volume}{76}}, \bibinfo{pages}{483}
  (\bibinfo{year}{1996}).

\bibitem{SW58}
\bibinfo{author}{\bibfnamefont{J.~C.} \bibnamefont{Slonczewski}}
  \bibnamefont{and} \bibinfo{author}{\bibfnamefont{P.~R.} \bibnamefont{Weiss}},
  \bibinfo{journal}{Phys. Rev.} \textbf{\bibinfo{volume}{109}},
  \bibinfo{pages}{272} (\bibinfo{year}{1958}).

\bibitem{Gon99}
\bibinfo{author}{\bibfnamefont{J.}~\bibnamefont{Gonz\'alez}},
  \bibinfo{author}{\bibfnamefont{F.}~\bibnamefont{Guinea}}, \bibnamefont{and}
  \bibinfo{author}{\bibfnamefont{M.~A.~H.} \bibnamefont{Vozmediano}},
  \bibinfo{journal}{\prb} \textbf{\bibinfo{volume}{59}}, \bibinfo{pages}{R2474}
  (\bibinfo{year}{1999}).

\bibitem{Khv01}
\bibinfo{author}{\bibfnamefont{D.~V.} \bibnamefont{Khveshchenko}},
  \bibinfo{journal}{\prl} \textbf{\bibinfo{volume}{87}},
  \bibinfo{pages}{246802} (\bibinfo{year}{2001}).

\bibitem{Khv01b}
\bibinfo{author}{\bibfnamefont{D.~V.} \bibnamefont{Khveshchenko}},
  \bibinfo{journal}{\prl} \textbf{\bibinfo{volume}{87}},
  \bibinfo{pages}{206401} (\bibinfo{year}{2001}).

\bibitem{CMW96}
\bibinfo{author}{\bibfnamefont{C.}~\bibnamefont{de~C.~Chamon}},
  \bibinfo{author}{\bibfnamefont{C.}~\bibnamefont{Mudry}}, \bibnamefont{and}
  \bibinfo{author}{\bibfnamefont{X.-G.} \bibnamefont{Wen}},
  \bibinfo{journal}{\prb} \textbf{\bibinfo{volume}{53}}, \bibinfo{pages}{R7638}
  (\bibinfo{year}{1996}).

\bibitem{Cetal97}
\bibinfo{author}{\bibfnamefont{H.~E.} \bibnamefont{Castillo}},
  \bibinfo{author}{\bibfnamefont{C.}~\bibnamefont{de~C.~Chamon}},
  \bibinfo{author}{\bibfnamefont{E.}~\bibnamefont{Fradkin}},
  \bibinfo{author}{\bibfnamefont{P.~M.} \bibnamefont{Goldbart}},
  \bibnamefont{and} \bibinfo{author}{\bibfnamefont{C.}~\bibnamefont{Mudry}},
  \bibinfo{journal}{\prb} \textbf{\bibinfo{volume}{56}}, \bibinfo{pages}{10668}
  (\bibinfo{year}{1997}).

\bibitem{HD02}
\bibinfo{author}{\bibfnamefont{B.}~\bibnamefont{Horovitz}} \bibnamefont{and}
  \bibinfo{author}{\bibfnamefont{P.~L.} \bibnamefont{Doussal}},
  \bibinfo{journal}{\prb} \textbf{\bibinfo{volume}{65}},
  \bibinfo{pages}{125323} (\bibinfo{year}{2002}).

\bibitem{G98}
\bibinfo{author}{\bibfnamefont{F.}~\bibnamefont{Guinea}},
  \bibinfo{journal}{\prb} \textbf{\bibinfo{volume}{58}}, \bibinfo{pages}{6622}
  (\bibinfo{year}{1998}).

\bibitem{SGV03}
\bibinfo{author}{\bibfnamefont{T.}~\bibnamefont{Stauber}},
  \bibinfo{author}{\bibfnamefont{F.}~\bibnamefont{Guinea}}, \bibnamefont{and}
  \bibinfo{author}{\bibfnamefont{M.~A.~H.} \bibnamefont{Vozmediano}}
  (\bibinfo{year}{2003}), \eprint{{\tt cond-mat/0311016}}.

\bibitem{VLG02}
\bibinfo{author}{\bibfnamefont{M.~A.~H.} \bibnamefont{Vozmediano}},
  \bibinfo{author}{\bibfnamefont{M.~P.} \bibnamefont{L\'opez-Sancho}},
  \bibnamefont{and} \bibinfo{author}{\bibfnamefont{F.}~\bibnamefont{Guinea}},
  \bibinfo{journal}{Phys. Rev. Lett.} \textbf{\bibinfo{volume}{89}},
  \bibinfo{pages}{166401} (\bibinfo{year}{2002}).

\bibitem{VLG03}
\bibinfo{author}{\bibfnamefont{M.~A.~H.} \bibnamefont{Vozmediano}},
  \bibinfo{author}{\bibfnamefont{M.~P.} \bibnamefont{L\'opez-Sancho}},
  \bibnamefont{and} \bibinfo{author}{\bibfnamefont{F.}~\bibnamefont{Guinea}},
  \bibinfo{journal}{Phys. Rev. B} \textbf{\bibinfo{volume}{68}},
  \bibinfo{pages}{195122} (\bibinfo{year}{2003}).

\bibitem{LFSG94}
\bibinfo{author}{\bibfnamefont{A.~W.~W.} \bibnamefont{Ludwig}},
  \bibinfo{author}{\bibfnamefont{M.~P.~A.} \bibnamefont{Fisher}},
  \bibinfo{author}{\bibfnamefont{R.}~\bibnamefont{Shankar}}, \bibnamefont{and}
  \bibinfo{author}{\bibfnamefont{G.}~\bibnamefont{Grinstein}},
  \bibinfo{journal}{Phys. Rev. B} \textbf{\bibinfo{volume}{50}},
  \bibinfo{pages}{7526} (\bibinfo{year}{1994}).

\bibitem{WS00}
\bibinfo{author}{\bibfnamefont{K.}~\bibnamefont{Wakayabashi}} \bibnamefont{and}
  \bibinfo{author}{\bibfnamefont{M.}~\bibnamefont{Sigrist}},
  \bibinfo{journal}{\prl} \textbf{\bibinfo{volume}{84}}, \bibinfo{pages}{3390}
  (\bibinfo{year}{2000}).

\bibitem{W01}
\bibinfo{author}{\bibfnamefont{K.}~\bibnamefont{Wakayabashi}},
  \bibinfo{journal}{\prb} \textbf{\bibinfo{volume}{64}},
  \bibinfo{pages}{125428} (\bibinfo{year}{2001}).

\bibitem{OS91}
\bibinfo{author}{\bibfnamefont{A.~A.} \bibnamefont{Ovchinnikov}}
  \bibnamefont{and} \bibinfo{author}{\bibfnamefont{I.~L.}
  \bibnamefont{Shamovsky}}, \bibinfo{journal}{Journ. of. Mol. Struc.
  (Theochem)} \textbf{\bibinfo{volume}{251}}, \bibinfo{pages}{133}
  (\bibinfo{year}{1991}).

\bibitem{GLSV04}
\bibinfo{author}{\bibfnamefont{F.}~\bibnamefont{Guinea}},
  \bibinfo{author}{\bibfnamefont{M.~P.} \bibnamefont{L\'opez-Sancho}},
  \bibinfo{author}{\bibfnamefont{T.}~\bibnamefont{Stauber}}, \bibnamefont{and}
  \bibinfo{author}{\bibfnamefont{M.~A.~H.} \bibnamefont{Vozmediano}}
  \eprint{{\tt Work in progress}}.

\bibitem{Vetal99}
\bibinfo{author}{\bibfnamefont{D.}~\bibnamefont{Vollhardt}},
  \bibinfo{author}{\bibfnamefont{N.}~\bibnamefont{Bl\"umer}},
  \bibinfo{author}{\bibfnamefont{K.}~\bibnamefont{Held}},
  \bibinfo{author}{\bibfnamefont{M.}~\bibnamefont{Kollar}},
  \bibinfo{author}{\bibfnamefont{J.}~\bibnamefont{Schlipf}},
  \bibinfo{author}{\bibfnamefont{M.}~\bibnamefont{Ulmke}}, \bibnamefont{and}
  \bibinfo{author}{\bibfnamefont{J.}~\bibnamefont{Wahle}}, in
  \emph{\bibinfo{booktitle}{Advances in Solid State Physics}}
  (\bibinfo{publisher}{Wieweg}, \bibinfo{year}{1999}).

\end{thebibliography}
\end{document}